\documentclass[10pt,letterpaper,twocolumn]{article}

\usepackage[margin=1in,columnsep=0.23in]{geometry}
\usepackage[american]{babel}

\usepackage{natbib}
\usepackage{mathtools}
\usepackage{booktabs}
\usepackage{tikz}
\usepackage{subcaption}
\usepackage{float}
\usepackage{amsthm}
\usepackage{amsfonts}
\usepackage{xfrac}
\usepackage{amssymb}
\usepackage{enumerate}
\usepackage{caption}
\usepackage{stmaryrd}
\usepackage{xspace}
\usepackage{environ}
\usepackage{tabularx}

\newtheorem{theorem}{Theorem}

\newtheorem{proposition}[theorem]{Proposition}

\newtheorem{hypothesis}[theorem]{Hypothesis}

\usetikzlibrary{decorations.pathreplacing}
\usetikzlibrary{arrows}

\newcommand{\be}{\begin{eqnarray}}
\newcommand{\ee}{\end{eqnarray}}
\newcommand{\bes}{\begin{eqnarray*}}
\newcommand{\ees}{\end{eqnarray*}}

\newcommand{\cC}{\mathcal{C}}

\newcommand{\cP}{\mathcal{P}}

\usepackage{xspace}
\newcommand{\BNSL}{\textsc{BNSL}\xspace}

\long\def\comment#1{}

\title{Quantum Speedups for Bayesian Network Structure Learning}

\author{
	\fontsize{10pt}{10pt} \textbf{Juha Harviainen}\\[-0.06cm]
	\fontsize{10pt}{10pt} \textrm{University of Helsinki}\\[-0.06cm]
	\fontsize{10pt}{10pt} \texttt{juha.harviainen@helsinki.fi}
	\and
	\fontsize{10pt}{10pt} \textbf{Kseniya Rychkova}\\[-0.06cm]
	\fontsize{10pt}{10pt} \textrm{University of Queensland}\\[-0.06cm]
	\fontsize{10pt}{10pt} \texttt{ksusha.rychkova@gmail.com}
	\and
	\fontsize{10pt}{10pt} \textbf{Mikko Koivisto}\\[-0.06cm]
	\fontsize{10pt}{10pt} \textrm{University of Helsinki}\\[-0.06cm]
	\fontsize{10pt}{10pt} \texttt{mikko.koivisto@helsinki.fi}
}
\date{}

\begin{document}
\maketitle

\begin{abstract}
The Bayesian network structure learning (BNSL) problem asks for a directed acyclic graph that maximizes a given score function. For networks with $n$ nodes, the fastest known algorithms run in time~$O(2^n n^2)$ in the worst case, with no improvement in the asymptotic bound for two decades. 
Inspired by recent advances in quantum computing, we ask whether BNSL admits a polynomial quantum speedup, that is, whether the problem can be solved by a quantum algorithm in time $O(c^n)$ for some constant $c$ less than $2$. 
We answer the question in the affirmative by giving two algorithms   
achieving $c \leq 1.817$ and $c \leq 1.982$ assuming the number of potential parent sets is, respectively, subexponential and $O(1.453^n)$. Both algorithms assume the availability of a quantum random access memory.
We also prove that one presumably cannot lower the base $2$ for any classical algorithm, as that would refute the strong exponential time hypothesis.  
\end{abstract}

\section{Introduction}

In the score-and-search approach to structure learning in Bayesian networks, one specifies a score function to be maximized over all possible directed acyclic graphs (DAGs) on a given node set. Common score functions---such as BDeu, BGe, BIC, fNML, or qNML---are decomposable: the score of a DAG is obtained by summing up the local scores of each node. The local score expresses how well the given parent set for a node fits the observed data, prior knowledge or constraints, and the adopted measures of learning success. See the textbook of \cite{Koller09} for other approaches and the survey of \citeauthor{Kitson23}~[\citeyear[Sec.~4.1]{Kitson23}] for descriptions of the scores.

Decomposability motivates studying a more abstract problem formulation, in which the local scores are treated as the input, effectively ignoring that they originate from a particular scoring metric and observed data. 
This optimization problem, known as \emph{Bayesian network structure learning} (BNSL),  
can be solved by dynamic programming over node subsets in time $O(2^n n^2)$, thus nearly linearly in the input size \citep{Ott04,Singh05,Silander06}. But what if the input consists of significantly fewer local scores, e.g., each node can have at most some constant number of parents---a case relevant in practice? Unfortunately, essentially no faster algorithms are known, the base of the exponential bound being stuck at~$2$. In fact, the problem is NP-hard already if the maximum indegree of the DAG is set to $2$ \citep{Chickering95NP-hard}. That said, there have been significant advances in heuristic algorithms, which may run fast for many practical instances \citep{Yuan13,Bartlett17}, 
as well as in parameterized algorithms, which admit improved worst-case time bounds for restricted problem variants (see \cite{Gruttemeier22} and references therein).  

Here, we ask whether quantum algorithms can beat the known exponential-time classical algorithms for BNSL. Quantum algorithms differ from classical ones in that they can harness quantum effects, such as superposition and entanglement. 
Typically (but not always) a quantum speedup is obtained by representing the problem in an appropriate way and then invoking a routine known as \emph{quantum search} or \emph{Grover's algorithm} \citep{Grover96}. Given a black-box mapping $f:\{1, 2, \dots, m\} \rightarrow \{0,1\}$, this routine only requires $O(\sqrt{m/k})$ evaluations of $f$ to find, with high probability, an element that maps to $1$, supposing there are $k$ such elements; the expected number of evaluations required by any classical algorithm is linear in $m/k$. Several problems are known to admit a quadratic quantum speedup in relation to the best known classical algorithms, examples ranging from the satisfiability problem \citep{Dantsin05} to learning linear classifiers \citep{Kapoor16,Roget22} and to reinforcement learning \citep{Dunjko16}. 

While there exist quantum \emph{approximation} algorithms for BNSL \citep{OGorman04,Soloviev23}, apparently, for \emph{exact} BNSL no quantum speedup was known before the present work. The main challenge is that the best classical algorithms already are significantly faster than exhaustive search over the super-exponentially many DAGs. The dynamic programming algorithms resemble the Bellman--Held--Karp algorithm \citep{Held61,Bellman62} for the traveling salesman problem and related ``permutation problems'' \citep{Koivisto10,Bodlaender12}, for which a quantum speedup was discovered only relatively recently \citep{Ambainis19}.

Inspired by the results of Ambainis et al., we will show that BNSL admits a quantum algorithm running in time $O(1.817^n F)$, where $F \leq n 2^{n-1}$ is the number of local scores given as input. This gives a polynomial speedup as long as $F$ grows subexponentially or very moderately exponentially in $n$. To give a polynomial speedup also when $F$ grows more rapidly, we present another, rather different algorithm: we make use of a construction previously given for trading space for time in a broad class of permutation problems \citep{Koivisto10}, including BNSL \citep{Parviainen13}. We give a quantum algorithm running in time $O(1.982^n)$, provided that $F = O(1.453^n)$. Both algorithms require a \emph{quantum random access memory} (QRAM) \citep{Giovannetti08}, of which experimental implementations do not yet exist.

Could the base $2$ be lowered also for a classical algorithm? Before the present work, the only evidence against has been the lack of progress in faster algorithms. Curiously, for a problem variant that ask for a sum over DAGs, the base of the exponential time bound was recently lowered from $3$ \citep{Tian09} to $2.985$ \citep{Koivisto20}. We will show that for the maximization variant, similar improvement presumably is not possible: we prove that it would refute the \emph{strong exponential time hypothesis} (SETH) and thus give a way to solve the CNF-SAT problem on $n$ variables in time $O(c^n)$ with some constant $c < 2$. 

The rest of this paper is organized as follows. Section~\ref{se:pr} introduces more formally the setup, namely the BNSL problem, the quantum search routine, and QRAM. Section~\ref{se:lo} gives our first algorithm and Section~\ref{se:po} the second. The connection to SETH is presented in Section~\ref{se:seth}. In Section~\ref{se:cm} we discuss some open problems and the role of QRAM. 

\section{Preliminaries}
\label{se:pr}

This section introduces the main ingredients needed in later parts of the paper. 

\subsection{Graphs and orders}

Let $N$ be a finite set and $R \subseteq N \times N$. We let
\bes
	R_i := \{ j : ji \in R \}
\ees
denote the \emph{parent set} (i.e., direct predecessors) of $i$ in $R$. 

If $R$ is \emph{acyclic}, i.e., there are no elements $i_1, i_2, \ldots, i_k$ such that $i_1 = i_k$ and $i_t i_{t+1} \in R$ for all $t = 1, 2,\ldots, k-1$, then~$(N, R)$ is a \emph{directed acyclic graph} (DAG). 

We call $R$ a \emph {partial order} on $N$ if it is irreflexive and transitive, and a \emph{linear order} on $N$ if it is, in addition, total (aka strongly connected). A linear order $L$ on $N$ is a \emph{linear extension} of $R$ if $L \supseteq R$, i.e., $L_i \supseteq R_i$ for all $i \in N$.

These concepts are illustrated in Figure~\ref{fig:order}. 

\begin{figure*}[!t]
	    \centering
	    \small
	    \begin{subfigure}[t]{0.32\textwidth}
			\centering
		    \begin{tikzpicture}
	    	    \node[shape=circle,draw] (N2) at (0,0) {2};
		        \node[shape=circle,draw] (N3) at (0,-1.5) {3};
		        \node[shape=circle,draw] (N6) at (0,-3) {6};
		        \node[shape=circle,draw] (N1) at (-1.5,-0.75) {1};
		        \node[shape=circle,draw] (N5) at (-1.5,-2.25) {5};
		        \node[shape=circle,draw] (N4) at (1.5,0) {4};
		        \node[shape=circle,draw] (N7) at (1.5,-1.5) {7};
		        \node[shape=circle,draw] (N8) at (1.5,-3) {8};
		        \path [-triangle 45] (N2) edge node[] {} (N1);
		        \path [-triangle 45] (N3) edge node[] {} (N1);
		        \path [-triangle 45] (N3) edge node[] {} (N5);
		        \path [-triangle 45] (N6) edge node[] {} (N5);
		        \path [-triangle 45] (N7) edge node[] {} (N5);
		        \path [-triangle 45] (N7) edge node[] {} (N3);
		        \path [-triangle 45] (N3) edge node[] {} (N2);
		        \path [-triangle 45] (N4) edge node[] {} (N2);
		        \path [-triangle 45] (N4) edge node[] {} (N3);
		        \path [-triangle 45] (N8) edge node[] {} (N6);
		    \end{tikzpicture}
		    \caption{DAG $A$}
		    \label{fig:dag}
		\end{subfigure}
		\begin{subfigure}[t]{0.32\textwidth}
			\centering
		    \begin{tikzpicture}
		        \node[shape=circle,draw] (N2) at (0,0) {2};
		        \node[shape=circle,draw] (N3) at (0,-1.5) {3};
		        \node[shape=circle,draw] (N6) at (0,-3) {6};
		        \node[shape=circle,draw] (N1) at (-1.5,-0.75) {1};
		        \node[shape=circle,draw] (N5) at (-1.5,-2.25) {5};
		        \node[shape=circle,draw] (N4) at (1.5,0) {4};
		        \node[shape=circle,draw] (N7) at (1.5,-1.5) {7};
		        \node[shape=circle,draw] (N8) at (1.5,-3) {8};
		        \path [-triangle 45] (N4) edge node[] {} (N7);
		        \path [-triangle 45] (N7) edge node[] {} (N8);
		        \path [-triangle 45] (N8) edge node[] {} (N6);
		        \path [-triangle 45] (N6) edge node[] {} (N3);
		        \path [-triangle 45] (N3) edge node[] {} (N2);
		        \path [-triangle 45] (N2) edge node[] {} (N1);
		        \path [-triangle 45] (N1) edge node[] {} (N5);
		    \end{tikzpicture}
		    \caption{Linear order $L$}
		    \label{fig:linear}
		\end{subfigure}
		\begin{subfigure}[t]{0.32\textwidth}
			\centering
		    \begin{tikzpicture}
		        \node[shape=circle,draw] (N1) at (0,0) {1};
		        \node[shape=circle,draw] (N2) at (0,-1) {2};
		        \node[shape=circle,draw] (N5) at (0,-2) {5};
		        \node[shape=circle,draw] (N6) at (0,-3) {6};
		        \node[shape=circle,draw] (N3) at (2,0) {3};
		        \node[shape=circle,draw] (N4) at (2,-1) {4};
		        \node[shape=circle,draw] (N7) at (2,-2) {7};
		        \node[shape=circle,draw] (N8) at (2,-3) {8};
		        \path [-triangle 45] (N3) edge node[] {} (N1);
		        \path [-triangle 45] (N3) edge node[] {} (N2);
		        \path [-triangle 45] (N4) edge node[] {} (N1);
		        \path [-triangle 45] (N4) edge node[] {} (N2);
		        \path [-triangle 45] (N7) edge node[] {} (N5);
		        \path [-triangle 45] (N7) edge node[] {} (N6);
		        \path [-triangle 45] (N8) edge node[] {} (N5);
		        \path [-triangle 45] (N8) edge node[] {} (N6);
		    \end{tikzpicture}
		    \caption{Partial order $P$}
		    \label{fig:partial}
		\end{subfigure}
	\caption{Examples of a DAG, a linear order, and a partial order on the node set $\{1, 2, \ldots, 8\}$. We have $A \subseteq L$ and $P \subseteq L$; for example, $A_6 = \{8\}$, $L_6 = \{4, 7, 8\}$, and $P_6 = \{7, 8\}$. For the linear order, shown is its transitive reduction, i.e., only the edges necessary for determining the relation uniquely using transitivity. The partial order $P$ is a member of the set of parallel bucket orders described in Section~\ref{se:po} (with $k = 4$).}
	\label{fig:order}
\end{figure*}

\subsection{The BNSL Problem}

Given a \emph{node set} $N$ of size $n$ and a \emph{local score} $s_i(J)$ for each node $i \in N$ and node subset $J \subseteq N \setminus \{i\}$, the BNSL problem is to find a DAG $(N, A)$ that maximizes the score 
\bes
	s(A) := \sum_{i \in N} s_i(A_i)\,,
\ees
which measures how well the DAG fits the prior assumptions and the data.
Here we identify the DAG with its \emph{arc set} $A$, the node set $N$ being fixed. Recall that $A_i$ denotes the parent set of $i$. Since our algorithms work for any decomposable score, we do not specify the used score; nevertheless descriptions of commonly used scores can be found on the survey of \citeauthor{Kitson23}~[\citeyear[Sec.~4.1]{Kitson23}].

Our interest is in instances in which most local scores equal~$-\infty$ and are not given as explicit input. Accordingly, for each node $i$ we are given a collection of potential parent sets $\cC_i$, the size of which can be substantially smaller than~$2^{n-1}$. The local scores $s_i(J)$ are only given for $J \in \cC_i$. We let 
\bes
	F := \sum_{i\in N}|\cC_i| 
\ees	
denote the total size of the input.

In practice, potential parent sets are obtained using several ideas and combinations thereof. One is to include in $\cC_i$ only sets that are contained in a relatively small set of candidate parents. Another idea is to only include sets whose cardinality does not exceed some given upper bound. A third technique is to exclude sets $J$ for which there is a subset~$J' \subseteq J$ with an equal or better local score, $s_i(J') \geq s_i(J)$; while this simple pruning rule may require computing the local scores for a large number of sets, more sophisticated analytic score bounds can also exclude sets without computing their scores \citep{Correia20}.
Importantly, all these procedures result in collections $\cC_i$ that are \emph{closed under inclusion}, that, if $J \in \cC_i$ and $J' \subseteq J$, then $J' \in \cC_i$. 

\subsection{Quantum Circuits and QRAM}

Quantum computation can be modeled by a quantum circuit that takes as input $\ell$ qubits representing the system's initial state, then transforms the state by reversible quantum logic gates, until the final state of some $l\leq \ell$ qubits of interest is measured and an $l$-bit output is obtained. The additional power of quantum circuits in comparison to classical boolean circuits stems from the fact that $\ell$ qubits can represent a superposition (i.e., a linear combination) of all the $2^\ell$ possible $\ell$-bit vectors. 
The coefficients (i.e., amplitudes) encode a probability distribution over the possible vectors, a measurement returning the corresponding random variable. A quantum algorithm is thus a randomized algorithm. It has \emph{bounded error} if, for all problem instances, the ouput is correct with probability at least $2/3$. 

For classical algorithms, the boolean circuit model can yield pessimistic complexity bounds---for more practical settings, one assumes a random access memory (RAM).\footnote{An algorithm that runs in time $T$ using a RAM can be simulated by a boolean circuit of size 
$T^2 \mathrm{polylog}(T)$ \citep{Cook73,Pippenger79}. 
} Similarly, broader applicability of quantum computation is believed to require an equivalent \emph {quantum RAM} (QRAM) \citep{Giovannetti08}. 
Importantly, QRAM enables invoking any time-$T$ classical algorithm that uses RAM as a $O(T)$-time subroutine in a quantum algorithm.

\subsection{Quantum Search}

Grover's algorithm \citep{Grover96}, also known as quantum search, is a celebrated generic algorithm for finding a needle in a haystack. 
As described in the Introduction, it gives a quadratic speedup in relation to classical algorithms. 
We will make use of the following powerful extension:
\begin{theorem}[\cite{Durr96}, \cite{Ambainis19}]\label{thm:grover}
Suppose $f(x)$ is an integer computable for any given $x \in \{1,2,\ldots,m\}$ by a bounded-error quantum algorithm in time $T$. 
Then there is a bounded-error quantum algorithm that computes $\max_{x=1}^m f(x)$ in time $O(T \sqrt{m} \log m)$.  
\end{theorem}

This result allows us to apply quantum search (i) in a maximization problem and (ii) recursively with only a negligible computational overhead.

\section{Finding a linear order}
\label{se:lo}

Various NP-hard graph problems can be viewed as finding an optimal node ordering. For our purposes it is convenient to consider the problem of computing 
\be \label{eq:lo}
	\max_{L} \sum_{i \in N} f(L_i, i)\,,
\ee
where the function $f$ depends on the problem input and the maximization is over all linear orders $L$ on~$N$. 

\cite{Ambainis19} gave a quantum algorithm for any problem of that form: 
\begin{theorem}[{\citet[Cor.~3.1]{Ambainis19}}]\label{thm:1817}
The problem (\refeq{eq:lo}) admits a bounded-error quantum algorithm that runs in time $O(1.817^n T)$, assuming $f$ can be evaluated in time $T$. 
\end{theorem}

It is easy to see---and well known \citep[Eq.~(9)]{Cooper92}---that BNSL can be written in the above form by putting 
\bes
	f(L_i, i) := \max_{J \subseteq L_i :  J \in \cC_i} s_i(J)
\ees
Indeed, if $A$ is an optimal DAG and $L$ a topological ordering of its nodes, the score $s(A)$ is obtained as $\sum_{i \in  N} f(L_i, i)$.  

Since $f(L_i, i)$ can be computed in time $O(|\cC_i|n)$ by a linear scan over the potential parent sets, we have got a quantum algorithm that solves BNSL in time $O(1.817^n F)$. To omit factors polynomial in $n$ in the asymptotic bound, we used the fact that the constant base $1.817$ of Theorem~\ref{thm:1817} was originally obtained by rounding up a strictly smaller constant.  

But we can do better. We simply replace the classical linear scan by quantum search:  
\begin{theorem}
BNSL admits a bounded-error quantum algorithm that runs in time $O(1.817^n \sqrt{F})$. 
\end{theorem}

If $F$ grows subexponentially in $n$, the bound can be simplified to $O(1.817^n)$. 
On the other hand, the base of the exponential exceeds the base $2$ of the fastest classical algorithms already if $F = \Omega(1.212^n)$. 
In the next section, we give a different quantum algorithm that beats the known classical algorithms as long as $F = O(1.453^n)$. 

\section{Covering by Partial Orders}
\label{se:po}

\cite{Koivisto10} presented the following approach to a broad class of permutation problems, including ones of the form (\refeq{eq:lo}). Let $\cP$ be a set of partial orders on $N$ such that every linear order on $N$ is an extension of at least one member in $\cP$; we call $\cP$ simply a \emph{cover} on $N$. Now, for any function $f$ of linear orders on $L$, we have  
\bes
	\max_L f(L) = \max_{P \in \cP} \max_{L \supseteq P} f(L)
\ees
where the first maximization is over all linear orders on $N$. One example is when $\cP = \{\emptyset\}$, rendering the outer maximization trivial. Another extreme case is when $\cP$ consists of all linear orders on $N$, rendering the inner maximization trivial. In general, we have decomposed the original problem into $|\cP|$ subproblems, each constrained by a different partial order. 

In particular, we can write the BNSL problem as 
\bes
	\max_{P \in \cP} g(P)\,,
\ees
with the subproblems   
\be \label{eq:po}
	g(P) : = \max_{L \supseteq P} \sum_{i \in N} \max_{J \subseteq L_i :  J \in \cC_i} s_i(J)\,.
\ee
\cite{Parviainen13} solved the subproblem by dynamic programming over the downsets of the partial order~$P$. A \emph{downset} is a subset
$S \subseteq N$ that is closed under the relation, i.e., if $i \in S$ and $ji \in P$, then $j \in S$. 
\begin{proposition}[{\citet[Theorem~16]{Parviainen13}}]
Suppose each $\cC_i$ is closed under inclusion. 
Then the subproblem (\refeq{eq:po}) admits an algorithm that runs in time $O(D n^2 +F n)$, where $D$ is the number of downsets of $P$. 
\end{proposition}

We are now ready to apply quantum search over the cover~$\cP$. Combining Theorem~\ref{thm:grover} with the above result for the subproblem gives us a quantum algorithm for BNSL: 
\begin{proposition}
Let $\cP$ be a cover on $N$, each $P \in \cP$ having $O(D)$ downsets. 
Suppose each $\cC_i$ is closed under inclusion and $\sum_i |\cC_i| = O(D)$.   
Then BNSL admits a bounded-error quantum algorithm that runs in time 
$O(D n^2 |\cP|^{1/2} \log |\cP|)$. 
\end{proposition}

For simplicity, we here restricted the sizes of the sets~$\cC_i$ so that the running time for the subproblem is dominated by the number of downsets; this restriction will ease our further running time analysis, but is not crucial for the correctness of the algorithm. 

Our goal is next to show that, with an appropriate choice of the cover $\cP$, the running time is $O(c^n)$ for some constant~$c$ less than $2$. Ignoring lower-order terms, our task is to minimize the product $D |\cP|^{1/2}$.

Fortuitously, essentially the same task is already addressed by \cite{Koivisto10} in disguise: they aim at minimizing the space--time product, i.e., the product of the space complexity and the time complexity, which is given by~$D^2 |\cP|$, again ignoring lower-order terms. (Both the space and the time requirement of classical dynamic programming over downsets scale roughly as $D$.)

They give the following construction of what they call \emph{parallel bucket orders} (of length two). Suppose $n$ is divisible by an even natural number $k$, which is a design parameter. Partition $N$ arbitrarily into $n/k$ sets $S_1, S_2, \ldots, S_{n/k}$ of size $k$. Let $\cP$ consist of all partial orders on $N$ of the form~$R^1 \cup R^2 \cup \cdots \cup R^{n/k}$, where each $R^t$ is a partial order on $S_t$ obtained by splitting $S_t$ into two subsets of size~$k/2$ so that the elements in one set precede all other elements in the other set, i.e., $R^t = S' \times S''$ for some disjoint $S', S'' \subset S_t$ with $|S'| = |S''| = k/2$. See Fig.~\ref{fig:order} for an illustration. 
Different values of $k$ yield different space--time tradeoff. The product is minimized at $k = 26$, with the following numbers. 

\begin{proposition}[\cite{Koivisto10}]
Let $N$ be an $n$-element set, with $n$ divisible by $26$.  
There is a cover~$\cP$ on $N$ with $a^{n/26}$ members, each having $b^{n/26}$ downsets, where $a := \binom{26}{13}$ and $b := 2^{14}-1$.   
\end{proposition}
 
The $n$-th root of $D |\cP| ^{1/2}$ is given by 
\bes
	a^{1/52} \cdot b^{1/26} < 1.3645 \cdot 1.4525 < 1.9820 =: c\,.
\ees
Since we round up the base $c$, the bound $O(c^n)$ suppresses any factor that grows subexponentially in $n$, including factors that arise when $n$ is not divisible by $26$ and the construction is modified accordingly (we omit details). 

\begin{theorem}
BNSL admits a bounded-error quantum algorithm that runs in time $O(1.982^n)$, provided that each $\cC_i$ is closed under inclusion and 
$F = \sum_i |\cC_i| = O(1.453^n)$.
\end{theorem}

\section{Computational Hardness}\label{se:seth}

In this section, we show that no classical algorithm can solve \BNSL in time $O(c^n)$ with $c < 2$, assuming the following \emph{strong exponential time hypothesis} (SETH) \citep{Impagliazzo01a,Impagliazzo01b}.

\begin{figure*}[!t]
    \centering
    \begin{tikzpicture}
        \node[shape=rectangle,draw=black] (s1) at (0,0) {$\{1, 2, 4\}$};
        \node[shape=rectangle,draw=black] (s2) at (2,0) {$\{1, 2\}$};
        \node[shape=rectangle,draw=black] (s3) at (4,0) {$\{2, 3, 4\}$};
        \node[shape=rectangle,draw=black] (s4) at (6,0) {$\{2, 3\}$};
        \node[shape=rectangle,draw=black] (s5) at (8,0) {$\{3, 4\}$};
        \node[shape=rectangle,draw=black] (s6) at (10,0) {$\{1, 4\}$};

        \node[shape=circle,draw=black] (u1) at (2,2.5) {$1$};
        \node[shape=circle,draw=black] (u2) at (4,2.5) {$2$};
        \node[shape=circle,draw=black] (u3) at (6,2.5) {$3$};
        \node[shape=circle,draw=black] (u4) at (8,2.5) {$4$};
        
        \path [-triangle 45] (u1) edge (s1);
        \path [-triangle 45] (u1) edge (s2);
        \path [-triangle 45] (u3) edge (s3);
        \path [-triangle 45] (u3) edge (s4);
        \path [-triangle 45] (u3) edge (s5);
        \path [-triangle 45] (u1) edge (s6);
        
        \path [-triangle 45] (s6) edge (s5);
        \path [-triangle 45] (s5) edge (s4);
        \path [-triangle 45] (s4) edge (s3);
        \path [-triangle 45] (s3) edge (s2);
        \path [-triangle 45] (s2) edge (s1);

        \path [-triangle 45, draw] (s1) to[out=80,in=-100]++ (0.4,2.5) to[out=80,in=150] (u2);
        \path [-triangle 45, draw] (s1) to[out=90,in=-90]++ (0.0,2) to[out=90,in=150] (u4);
    \end{tikzpicture}
    \caption{An optimal structure in a reduction from a 3-\textsc{Hitting Set} instance with a universe of $4$ elements (circles) and a family of $6$ sets (rectangles) to \textsc{BNSL}.}\label{fig:simple}
\end{figure*}

\begin{hypothesis}[SETH]
    For any $\delta < 1$ there exists a number $k$ such that the \textsc{$k$-CNF-SAT} problem over $n$ variables cannot be solved in time $O\big(2^{\delta n}\big)$ by a classical algorithm. 
\end{hypothesis}

To connect the hardness of \BNSL to SETH, we will construct a reduction from the
$k$-\textsc{Hitting Set} problem: given a universe $U$ of size $n$ and a family $\mathcal{T}$ of subsets of $U$ with at most $k$ elements, is there a subset of $U$ of size $t$ that intersects all members of $\mathcal{T}$?

\begin{theorem}[\cite{Cygan16}]\label{thm:hs}
    If SETH holds, for any $\delta < 1$ there exists a number $k$ such that the \textsc{$k$-Hitting Set} problem over a universe of size $n$ cannot be solved in time $O\big(2^{\delta n}\big)$ by a classical algorithm. 
\end{theorem}

We state our result for \BNSL in a form that replaces the parameter $k$ above by the restriction that the input size is subexponential in the number of nodes. We leave it as an open problem to improve this to a polynomial bound. 
\begin{theorem}\label{thm:seth}
    If SETH holds, the \textsc{BNSL} problem over $n$ variables and $2^{o(n)}$ potential parent sets cannot be solved in time $O\big(2^{\delta n}\big)$ for any $\delta < 1$  by a classical algorithm. 
\end{theorem}
\begin{proof}
    Consider an instance $(U, \mathcal{T}, k, t)$ of the \textsc{$k$-Hitting Set} problem, where $U = \{u_1, u_2, \dots, u_n\}$ is the universe and $\mathcal{T} = \{T_1, T_2, \dots, T_m\}$ is a family of subsets of $U$ of size at most $k$. 

    We first give a simpler reduction that results in a \BNSL instance with $n + m$ nodes. Then, we continue by \emph{sparsifying} the obtained instance by merging some of the nodes, rendering the number of nodes independent of $m$. Finally, we show that solving that instance in time $O\big(2^{\delta n}\big)$ for any~$\delta < 1$ would break SETH.

    We construct a \BNSL instance where the nodes correspond to the $n$ elements of the universe $U$ and the $m$ subsets in the given family $\mathcal{T}$; we denote these nodes with the same symbols $u_i$ and $T_j$ for notational convenience.

    Define the following local scores (the rest being $-\infty$):
    \begin{itemize}
        \item[] $s_{u_i}(\emptyset) = 0$,
        \item[] $s_{u_i}(\{T_1\}) = 1$,
        \item[] $s_{T_j}(\{u_i, T_{j+1}\}) = 0$ if $u_i \in T_j$ and $j < m$,
        \item[] $s_{T_m}(\{u_i\}) = 0$ if $u_i \in T_m$.
    \end{itemize}

    Suppose that $H$ is a hitting set of $\mathcal{T}$. Then, the following parent set assignment is possible, that is, it yields a nonnegative score:
    \begin{itemize}
        \item[] $A_{u_i} = \emptyset$ if $u_i \in H$,
        \item[] $A_{u_i} = \{T_1\}$ if $u_i \not\in H$,
        \item[] $A_{T_j} = \{u_i, T_{j+1}\}$ for some $u_i \in H$ if $j < m$,
        \item[] $A_{T_m} = \{u_i\}$ for some $u_i \in H$.
    \end{itemize}
    Such a DAG attains a score $n - |H|$. An illustration is provided in Figure~\ref{fig:simple}.

	\begin{figure*}[!t]
		\centering
		\begin{tikzpicture}
			\node[shape=rectangle,draw=black] (s123) at (0.5,-1.5) {\begin{tabular}{c}
				$\{1, 2, 3\}$\\
				\midrule
				$\{1, 2\}$\\
				$\{2, 3\}$\\
			\end{tabular}};
			\node[shape=rectangle,draw=black] (s124) at (3.5,-1.5) {\begin{tabular}{c}
				$\{1, 2, 4\}$\\
				\midrule
				$\{1, 2\}$\\
				$\{1, 2, 4\}$\\
				$\{1, 4\}$\\
			\end{tabular}};
			\node[shape=rectangle,draw=black] (s134) at (6.5,-1.5) {\begin{tabular}{c}
				$\{1, 3, 4\}$\\
				\midrule
				$\{1, 4\}$\\
				$\{3, 4\}$\\
			\end{tabular}};
			\node[shape=rectangle,draw=black] (s234) at (9.5,-1.5) {\begin{tabular}{c}
				$\{2, 3, 4\}$\\
				\midrule
				$\{2, 3\}$\\
				$\{3, 4\}$\\
				$\{2, 3, 4\}$\\
			\end{tabular}};
	
			\node[shape=circle,draw=black] (u1) at (2,2) {$1$};
			\node[shape=circle,draw=black] (u2) at (4,2) {$2$};
			\node[shape=circle,draw=black] (u3) at (6,2) {$3$};
			\node[shape=circle,draw=black] (u4) at (8,2) {$4$};
			
			\path [-triangle 45] (u1) edge (s123);
			\path [-triangle 45] (u1) edge (s124);
			\path [-triangle 45] (u1) edge (s134);
			\path [-triangle 45] (u3) edge (s123);
			\path [-triangle 45] (u3) edge (s134);
			\path [-triangle 45] (u3) edge (s234);
			
			\path [-triangle 45] (s234) edge (s134);
			\path [-triangle 45] (s134) edge (s124);
			\path [-triangle 45] (s124) edge (s123);
	
			\path [-triangle 45, draw] (s123) to[out=80,in=-100]++ (0.4,3.0) to[out=80,in=150] (u2);
			\path [-triangle 45, draw] (s123) to[out=90,in=-90]++ (0.0,2.5) to[out=90,in=150] (u4);
		\end{tikzpicture}
		\caption{An optimal structure in a sparsified reduction from a 3-\textsc{Hitting Set} instance with a universe of $4$ elements and a family of $6$ sets to \textsc{BNSL}. Each set of the input is associated with at least one of the nodes, where the superset is written above the horizontal line and the associated subsets below it.}\label{fig:sparsified}
	\end{figure*}

	We claim that if $H$ is a minimum-size hitting set for $\mathcal{T}$, then no DAG can exceed the score $n - |H|$. First, note that any DAG $A$ with a nonnegative score corresponds to a hitting set
	\bes
		H_A \coloneqq \{u_1, u_2, \dots, u_n\} \cap \Bigg( \bigcup_{j=1}^m A_{T_j} \Bigg)
	\ees
	for $\mathcal{T}$, since the parent set of each $T_j$ must include a node $u_i$ with $u_i \in T_j$ by the definition of the local scores. Further, the nodes $u_i \in H_A$ cannot have any parents, since otherwise this would violate acyclicity: the only potential non-empty parent set of $u_i$ is $\{T_1\}$, but the DAG has to contain edges~$T_1 \leftarrow T_2 \leftarrow \dots \leftarrow T_j$ and $T_j \leftarrow u_i$ for some~$j$. Finally, if $u_i \not\in H_A$, then $u_i$ has no children and can pick any of its potential parent sets. In particular, its local score is maximized by choosing $\{T_1\}$ with score $1$. Thus, the \BNSL instance admits a solution with score $n - t$ if and only there is a hitting set of size $t$ for $\mathcal{T}$.
	
    The constructed \BNSL instance has $n + m$ nodes, which is too many to prove our theorem. We next sparsify the subset of $m$ nodes that correspond to the members in $\mathcal{T}$.

    We arbitrarily partition the universe $U$ into $p \coloneqq \lceil n^{1/(k+1)} \rceil$ sets of (almost) equal size, $U_1, U_2, \dots, U_{p}$, that is, their sizes~$|U_i|$ differ by at most $1$. For all $I \subseteq \{1, 2, \dots, p\}$, let 
\bes
	U_I \coloneqq \bigcup_{i \in I} U_i\,.
\ees	
Note that any $T \in \mathcal{T}$ is a subset of $U_I$ for some $I$ of size $k$. 

Instead of introducing a node for each $T_j$ in our \BNSL instance, we introduce a node for each $U_I$ with $|I| = k$. Label these sets arbitrarily by $T'_1, T'_2, \dots, T'_{m'}$ with~$m' = {\binom{p}{k}}$. For a subset $P$ of $U$ say that $P$ \emph{hits} $T'_j$ if $P \subseteq T'_j$ and~$P$ intersects all $T \in \mathcal{T}$ with $T \subseteq T'_j$.
	
Define the following local scores (and potential parent sets):
    \begin{itemize}
        \item[] $s_{u_i}(\emptyset) = 0$,
        \item[] $s_{u_i}(\{T'_1\}) = 1$,
        \item[] $s_{T'_j}(P \cup \{T'_{j+1}\}) = 0$ if $P$ hits $T'_j$ and $j < m'$,
        \item[] $s_{T'_{m'}}(P) = 0$ if $P$ hits $T'_{m'}$.
    \end{itemize}

    In other words, the new local scores ensure that the parent set of $T'_{j}$ hits all members of $\mathcal{T}$ that are its subsets.

As before, for any minimum-size hitting set $H$ of $\mathcal{T}$, 
the maximum score $n - |H|$ is attained by the following parent set assignments:
    \begin{itemize}
        \item[] $A_{u_i} = \emptyset$ if $u_i \in H$,
        \item[] $A_{u_i} = \{T_1\}$ if $u_i \not\in H$,
        \item[] $A_{T'_j} = (H \cap T'_j) \cup \{T'_{j+1}\}$ if $j < m'$,
        \item[] $A_{T'_{m'}} = H \cap T'_{m'}$.
    \end{itemize}
This is illustrated in Figure~\ref{fig:sparsified}.

The instance now contains $n + \binom{p}{k}$ nodes. Since~$p = \lceil n^{1/(k+1)}\rceil$, we have $\binom{p}{k} \le p^k = o(n)$. 	

To bound the number of potential parent sets, observe that each node $T'_j$ has at most $2^{|T'_j|}$ potential parent sets. Since $T'_j$ is a union of $k$ parts $U_i$, each part of size at most~$\lceil n/p \rceil \leq \lceil n^{k/(k+1)} \rceil = o(n)$, the number of potential parent sets of~$T'_j$ is at most $2^{k \lceil n/p \rceil} = 2^{o(n)}$. The total number of potential parent sets is thus bounded by 
\bes
	2 n + \binom{p}{k} 2^{o(n)} = 2^{o(n)}\,.
\ees

In summary, constructing the instance takes subexponential time, there are subexponentially many potential parent sets, and the number of nodes is asymptotically equivalent to $n$.

Assume now that SETH holds but any instance of \BNSL with $n'$ nodes could be solved in time $O(2^{\delta' n'})$ for some~$\delta' < 1$. Put $\delta \coloneqq (\delta'+1)/2 < 1$. By Theorem~\ref{thm:hs}, there exists a $k$ such that \textsc{$k$-Hitting Set} with a universe of size $n$ cannot be solved in time $O\big(2^{\delta n}\big)$. However, we showed that any such instance can be reduced to an instance of \BNSL on $n' = n + o(n)$ variables in time $2^{o(n)}$. By our assumption, we can solve it in time~$O(2^{\delta' \cdot (n+o(n))}) = O(2^{o(n)} 2^{\delta' n}) = O(2^{\delta n})$, which is a contradiction. 
\end{proof}

\section{Concluding remarks}
\label{se:cm}

We have shown quantum speedups for the problem of Bayesian network structure learning. Our two algorithms are built on rather sophisticated previous results: a quantum algorithm for a related problem \citep{Ambainis19} and a classical algorithm for the same problem \citep{Parviainen13}. On the other hand, the ways we employed these previous results are technically relatively simple. 
We also proved that similar speedups presumably are not possible for classical algorithms, suggesting that the \BNSL problem admits a ``quantum advantage.''\footnote{We use scare quotes because some researchers and practitioners reserve the term advantage for speedups that are exponential or experimentally demonstrated.}

An obvious question for further research is whether there are significantly faster quantum algorithms, e.g., ones with running time close to $O(2^{n/2})$ or others that yield a quantum speedup even when we do not bound the number of potential parent sets.  
Achieving the former target would most likely imply a new algorithm for the travelling salesman problem that beats current time bound of $O(1.728^n)$ \citep{Ambainis19}. 
The latter question assumes that the local scores are given implicitly, which is not an obstacle per se, as the local score of a given node and parent set can be computed efficiently from data for commonly used scoring functions.

Our algorithms may not have practical value in the near future. The speedup factor $(2/1.817)^n$ of our first algorithm achieves $10$ at $n \approx 24$ and $100$ at $n \approx 48$. However, the hidden subexponential factors are likely to favor the classical algorithms in practice even if the number of potential parent sets $F$ is small, say, cubic in $n$. 
Perhaps most importantly, our algorithms rely on QRAM, of which size is exponential in $n$. While different QRAM architectures have been proposed \citep{Giovannetti08,Park19}, there is no physical realization of the ideas yet. 
Currently we do not know whether the role of QRAM is critical for achieving any polynomial quantum speedup.

Regarding lower bounds for classical algorithms, there are several directions for future research. Theorem~\ref{thm:seth} does not rule out a faster algorithm when the maximum size of any potential parent set is bounded by a constant; the problem is solvable in polynomial time when the maximum size is one \citep{Chu65,Edmonds67}, but for larger upper bounds we only know that the problem is NP-hard \citep{Chickering95NP-hard}. One could also attempt to prove conditional lower bounds under some other established hypothesis not known to be implied by SETH such as the set cover conjecture \citep[p.~507]{Cygan15}.

\section*{Acknowledgements}

Research partially supported by the Research Council of Finland, grants 316771 and 351156.

\bibliography{paper}

\end{document}